\documentstyle[12pt,a4,epsf,rotate]{article}
\newcommand{\rcite}[1]{{\cite{#1}}}
\newcommand{\rref}[1]{{(\ref{#1})}}
\newcommand{\tref}[1]{{\ref{#1}}}
\newcommand{\rlabel}[1]{{\label{#1}}}
\newcommand{\rbibitem}[1]{\bibitem{#1}}
\newcommand{\dis}{\displaystyle}

\def\theequation{\arabic{section}.\arabic{equation}}

\newcommand{\eps}{\epsilon}

\newcommand{\beq}{\begin{equation}} 
\newcommand{\eeq}{\end{equation}}
\newcommand{\ba}{\begin{eqnarray}} 
\newcommand{\ea}{\end{eqnarray}}
\newcommand{\bd}{\begin{displaymath}}
\newcommand{\ed}{\end{displaymath}}
\newcommand{\br}{\beq\renewcommand{\arraystretch}{1.3}\begin{array}{l}}

\begin{document}
 
\begin{titlepage} 
\begin{flushright}
NORDITA 96/31 N,P\\ 
hep-ph/9606286\\
\end{flushright}
\vspace{0.5cm} \begin{center}
 
{ \LARGE \bf The mass of the scalar boson\\
\vspace{0.5cm}
beyond the large-$N_c$ limit}

\vspace*{1.5cm}
      {\bf E. Pallante\footnotemark
      \footnotetext{ email: pallante@nbivax.nbi.dk}} \\ 
{\footnotesize{ NORDITA, Blegdamsvej 17,
2100 Copenhagen, Denmark}} \\   \end{center}

\vspace*{1.0cm}

\begin{abstract}

Within the framework of the $1/N_c$ expansion of four-fermion interaction 
models, we analyse the next to leading $1/N_c$ corrections 
to the well known large-$N_c$ result $M_S=2M_Q$ where $M_S$ 
is the mass of the scalar boson and $M_Q$ is the 
constituent quark mass. The calculation is performed in the 
Extended Nambu-Jona Lasinio (ENJL) model which is suitable for 
describing low energy hadron properties.
We treat the model as fully non renormalizable and discuss the comparison 
with approaches based on the equivalence with renormalizable Yukawa type 
models.
We consider both the $G_V=0$ and the $G_V\neq 0$ cases with $n_f=2$ 
flavours and study the 
dependence upon the regularization scheme.
We find that pure next-to-leading $1/N_c$ corrections are large and 
negative, while a partially resummed treatment can induce positive and 
smaller corrections. A triplet-singlet states' splitting is observed.

\end{abstract} \vspace*{1.5cm} \begin{flushleft} 
NORDITA 96/31 N,P 
\end{flushleft} \end{titlepage} \newpage 

\section{\bf Introduction}
\setcounter{equation}{0}

The physical content of four-fermion interaction models has been 
extensively analysed in the past recent years. Within the $1/N$ 
expansion approach \rcite{oneoverN} for a general $U(N)$ symmetric
model, the equivalence {\em under certain assumptions} of 
four-fermion models with scalar four-fermion interactions
and Yukawa-type models has been investigated in
\rcite{ZJ,Hasenfratz}, while the consequences of imposing the so called
{\em compositness condition} have been derived by
\rcite{Shyzuya,others} and most recently by \rcite{Akama}.
In all the cases the renormalized ratio of boson and fermion masses 
plays a relevant role. In \rcite{ZJ} is shown that the ratio
goes to a fixed value due to the infrared freedom of the renormalizable 
Yukawa-type model and the assumption that the couplings are generic at 
the cut-off scale.
In addition a common trend of all the analyses seems to be the fact that 
the large-$N_c$ value of the mass ratio $M_S/M_Q=2$ gets a large
and negative next to leading $1/N_c$ correction for a realistic 
value $N_c=3$. This
suggests an asymptotic behaviour of the series where each finite order 
in $1/N_c$ fails to give a good estimate of the real value of the mass ratio
for useful values of $N_c$.

In this paper we address a calculation of the $1/N_c$ next to leading 
correction to the scalar boson mass based on a treatment 
of a four-fermion 
model {\em \`a la} Nambu-Jona Lasinio  which is alternative to the 
approaches formulated in \rcite{ZJ} and \rcite{Akama} and is of 
more immediate use in the derivation of hadron properties. 
In this case the model is treated as fully non renormalizable
(see \rcite{KLEVANSKY} and \rcite{BIJNENS} for reviews).

Effective constituent quark models {\em \`a la} Nambu-Jona Lasinio 
have been found to be successful in
reproducing the experimental values of the low-energy coupling constants
to $O(p^4)$ in Chiral Perturbation Theory (ChPt)
\rcite{ENJL}. Here the proper time regularization has been used.
Small dependence upon the regularization scheme has been also verified in 
\rcite{GRANDE,KLEVANSKY}.
Many of the couplings between resonances and pseudoscalar
mesons
have been  computed \rcite{ENJL,PRADES} and nicely compare with
the experiment,  as well as the vector and axial-vector masses
\rcite{ENJL,2point}. 

Large-$N_c$ two and three point functions have been derived in the fully 
fermionic language via the 
resummation of linear chains of constituent quark bubbles (sausage 
diagrams of the $\Phi^4$ theory) \rcite{2point,3point}.
In all the phenomenological results the explicit dependence upon the 
ultraviolet cut-off of the effective theory is kept treating the model
as fully non renormalizable and away from the infrared limit.
The large-$N_c$ calculation of the scalar two point function in the 
fully fermionic language and in the chiral limit reproduces 
a pole at $M_S = 2M_Q$ \rcite{2point}, where $M_Q$ is the constituent 
quark mass. With a typical value of $M_Q = 250\div 350$ MeV  
one has $M_S = 500\div 700$ MeV. 
The question arises if this pole has to be identified with a physical 
hadron state or it remains an artefact of the low energy model possibly 
related to the lack of confinement.
One remote possibility is that the eventual low lying scalar resonance 
has a very large width.

From the experimental point of view
a signal of a narrow scalar state around 750 MeV
is reported in \rcite{SVEC},
while the first  clear scalar resonances are the $a_0(983)$ and the
isosinglet $f_0(975)$ states. 
Their interpretation as an ordinary $q\bar{q}$ bound state is dubious
\rcite{BRAMON,ACHASOV,WEINSTEIN}.
Most recently a fit of the available data \rcite{TORN}
indicated that the $K\bar{K}$ component is large for both the $a_0(983)$ 
and $f_0(975)$ states.
It is then clear that the identification of the physical scalar 
states $a_0$ and 
$f_0$ would probably require the insertion of a mixing with exotic 
states (glueballs, $K\bar{K}$ bound states etc.) inside a low energy 
model. This is beyond the scope of this paper.

The present version of the ENJL model only allows for a scalar state 
which can be elementary or a $\bar{q}q$ composite state. 
Nonetheless we show that $1/N_c$ corrections to the large-$N_c$ 
value of the scalar mass can produce a splitting 
between the octet and the singlet scalar states.

In section \tref{NB} we first outline the model and make clear the 
main differences amongst the present approach and the approaches
in \rcite{ZJ} and \rcite{Akama} based on the equivalence of the
four-fermion interaction models with a renormalizable Yukawa-type 
model. Then we clarify the correspondence between the 
non-bosonized version of the ENJL model, where only fermion 
degrees of freedom are 
present, and its bosonized version which only contains the 
auxiliary boson fields once the fermions have been integrated out.
The appearance of overlapping divergences in the diagrams which give the 
$1/N_c$ n.t.l. corrections to the scalar two-point function
in the non-bosonized version can prevent from a simple 
and unambiguous calculation. We chose to compute them in the bosonized
version which gives a reliable and fully analytical approximation
of the exact result.
In section \tref{2} we derive the $1/N_c$ corrections to the scalar pole 
mass in the $G_V =0$ case, where only scalar and pseudoscalar meson 
fields are present in the bosonized action and with $n_f=2$ flavours. 
Here a subsection is 
dedicated to the treatment of leading divergences in this type of 
theories required by chiral invariance.
We also comment on different covariant regularization schemes.
In section \tref{3} we extend the model to the case $G_V\neq 0$, where 
also vector and axial-vector fields are present.
For both cases a numerical analysis is shown and the appearance of a 
mass splitting between the scalar singlet and the non-singlet is 
obtained.
We comment on numerical results and state our conclusions in section 
\tref{CONC}.

\section{\bf Bosonized versus non-bosonized version of the ENJL model.}

\setcounter{equation}{0}
\rlabel{NB}

The effective ENJL Lagrangian can be written as follows \rcite{ENJL}:
\beq
\rlabel{ENJL} 
{\cal L}_{ENJL} ={\cal L}_{QCD}^{\Lambda} + {\cal L}_{S,P}+{\cal 
L}_{V,A},
\eeq
where ${\cal L}_{S,P}$ and ${\cal L}_{V,A}$ are all the 
possible four-fermion lowest dimensional interactions allowed by chiral 
symmetry and leading in the $1/N_c$ expansion
\begin{eqnarray}
{\cal L}_{S,P}&=&
{8\pi^2 G_S(\Lambda )\over
N_c\Lambda^2}\sum_{a,b} 
(\bar{q}^a_Rq_L^b)(\bar{q}^b_Lq_R^a)={4\pi^2 G_S\over
N_c\Lambda^2}[(\bar{q}q)^2-(\bar{q}\gamma_5q)^2]  
\nonumber\\
{\cal L}_{V,A}&=&-{8\pi^2
G_V(\Lambda )
\over N_c\Lambda^2} \sum_{a,b}[(\bar{q}^a_L\gamma_\mu q_L^b)
(\bar{q}^b_L\gamma_\mu q_L^a)+(L\to R)] .
\end{eqnarray}
At this level the model contains only fermion d.o.f. and is written in 
terms of three independent parameters: $G_S$, 
$G_V$, and the physical cut-off $\Lambda$ of the effective interaction.
New extra parameters can be hidden in the cut-off procedure which is 
necessary in a non renormalizable model.
The couplings $G_S$ and $G_V$ are explicitly dependent upon the cut-off
and we have pulled out a factor $1/N_c$ so that they are $O(1)$ in the 
$1/N_c$ expansion; $a,b$ are flavour indices and a sum over colour d.o.f 
is implicit between brackets.
${\cal L}_{QCD}^{\Lambda}$ is the QCD Lagrangian in 
the presence of external sources and in the presence of a 
low energy cut-off where high frequency quark 
and gluon modes (i.e. with energy greater than $\Lambda$) have been 
integrated out.
The problem of the connection between QCD and this  Lagrangian
has been addressed in \rcite{ENJL}. 
The non renormalizable (by power counting) part of the Lagrangian 
\rref{ENJL} is in principle the first 
term of a double expansion in $1/N_c$, where $N_c$ is the number of 
colours, and in  $1/\Lambda^2$.

It is worth at this stage
to outline the main differences of our approach with recent 
analyses of four-fermion models based on their equivalence with 
renormalizable Yukawa-type models at least in the case $G_V=0$.
There are essentially two pictures explored alternative to the present 
one. A quite general RG equations analysis of the Gross-Neveu (GN) model
has been done by 
Zinn-Justin in \rcite{ZJ} (see also \rcite{Hasenfratz} for a 
numerical study of a NJL model on the same lines), 
while a NJL model has been studied in \rcite{Akama}. 
Here the consequences of imposing the 
{\em compositness condition} on the scalar field of the renormalizable 
Yukawa model as an additional constraint are analysed 
(see also refs. therein).

In the RGE analysis in \rcite{ZJ} the mass gap equation and the scalar 
propagator of the renormalizable generalized GN model in four dimensions
reduce to the ordinary GN ones in the 
infrared limit $\sigma ,p\ll\Lambda$, where $\sigma$ is the vacuum 
expectation value of the scalar field and $p$ is the typical 
four-momentum. Then in this case the equivalence of the 
four-fermion model with the generalized GN model is strictly valid in 
the {\em infrared domain} within the $1/N$ expansion.
The equivalence can be in principle spoiled beyond the $1/N$ 
expansion for small values of $N$. As also pointed out in \rcite{ZJ} 
this regime can be investigated by numerical studies of the four-fermion 
model in four dimensions and compared with analytic $\epsilon$ 
expansion of the renormalizable model.
The type of equivalence in \rcite{ZJ} also allows for the 
presence of higher dimensions 
operators in the four-fermion model which are irrelevant in $d<4$ 
dimensions. 

The {\em compositness condition} on the scalar field of a renormalizable
Yukawa model treated in \rcite{Akama} is a stronger extra constraint 
which guarantees 
the equivalence between a NJL model and a Yukawa one 
and which affects the RG flow of the renormalized couplings
of the Yukawa model. It spoils 
the renormalizability of the Yukawa model in four dimensions. 
One underlying difference between the approaches in \rcite{ZJ} and 
\rcite{Akama} is the fact that 
the compositness condition implies a particular value of the {\em bare} 
couplings at the cut-off scale while they are {\em naturally} assumed to be 
generic in the RG analysis of \rcite{ZJ}. 
The analysis in \rcite{Akama} also provides a prediction for the 
ratio of the boson and fermion masses at next-to-leading order in the 
$1/N$ expansion.
As will be also true in our case, $1/N$ next-to-leading corrections 
to the mass ratio are large and negative.

The main difference with our approach is that they keep the original 
four-fermion model at the infrared limit, which actually
corresponds to the limit $\Lambda\to\infty$ or equivalently $\sigma 
,p\ll\Lambda$.
What we do is to keep the model {\em away from the infrared domain}, 
which corresponds to being away from the limit $\sigma\ll\Lambda$
in the RGE analysis and in the solution of the mass gap equation.
In this case the four-fermion model stays as non renormalizable and $1/N_c$ 
corrections to the scalar mass can in principle be derived in the 
non-bosonized version where only constituent quarks appear.

\begin{figure}
\begin{center}
\leavevmode\epsfxsize=10cm\epsfysize=1.5cm\epsfbox{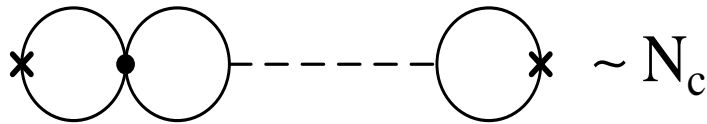}
\end{center}
\caption{ Two-point function in the non bosonized ENJL model in the 
large-$N_c$ limit   }
\label{fig1}
\end{figure}

A full two-point function in the large-$N_c$ limit is given by the 
infinite resummation of linear chains of quark bubbles 
with the insertion of a four-quark interaction vertex as shown in 
Fig.\tref{fig1}. It is of order $N_c$ (each fermion loop gives a 
factor $N_c$). Two and three-point functions have been derived in 
\rcite{2point,3point}.
Analogously a full n-point function in the large-$N_c$ limit is given by 
the one-constituent quark loop dressed by the insertion of two-point 
function legs attaching to the external sources.

\begin{figure}
\begin{center}
\leavevmode\epsfxsize=10cm\epsfysize=3cm\epsfbox{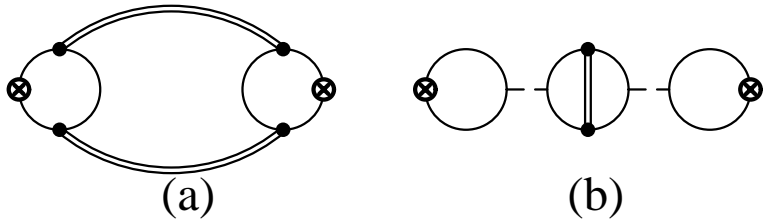}
\end{center}
\caption{Two-point function in the non bosonized ENJL model at 
next-to-leading order in $1/N_c$. 
Diagrams (a) are self-energy insertions. 
Diagrams (b) are vertex corrections. The double lines are linear chains
of constituent quark bubbles.}
\label{fig2}
\end{figure}
Next to leading in $1/N_c$ corrections are given in the diagrammatic 
language by all the possible insertions of one loop of chains of 
quark bubbles in the 
large-$N_c$ diagrams. They are of two types: self-energy insertions
(Fig.\tref{fig2}(a)) and vertex corrections (Fig.\tref{fig2}(b)).
Being a non renormalizable model the one loop correction implies a new 
divergence and thus a new counterterm which we keep the {\em new} cut-off 
$\tilde\Lambda$ of the loop.
The exact calculation of the next to leading in $1/N_c$ corrections
to the scalar two-point function (and in particular to its pole mass)
involves the one loop insertions of the type \tref{fig2}(a) and 
\tref{fig2}(b) in the large-$N_c$ scalar two-point function of 
Fig.\tref{fig1}. Defining the scalar two-point function as
$\Pi(q^2)=i\int d^4x~ e^{iqx} \langle 0 \vert T S(x) S(0)\vert 
0\rangle$,
where $S(x)\equiv -{1\over \sqrt{2}}\bar q(x) q(x)$
(we omit for simplicity flavour indices), the large-$N_c$
expression is given by \rcite{2point}
\beq
\Pi(Q^2)^{(N_c\to\infty)} =\bar\Pi(Q^2)
\sum_{n=0}^{\infty}\left(g_S\bar\Pi(Q^2)\right)^n=
{\bar \Pi(Q^2)\over 1 - g_S \bar\Pi(Q^2)} ,
\eeq
where
\beq
\dis \bar\Pi(Q^2)={1\over g_S} - (Q^2+(2M_Q)^2)Z_S(Q^2)
\eeq
is the bare fermion loop diagram in the mean-field
approximation and $Z_S(Q^2)$ is the scalar 
wave function renormalization constant which in the proper time 
regularization is given by
\beq
Z_S(Q^2)={N_c\over 16\pi^2}2 \int_0^1 ~d\alpha \Gamma \biggl ( 0,
{\alpha(1-\alpha)Q^2+M_Q^2\over\Lambda^2} \biggr ) ,
\eeq
\noindent with $\Gamma (0, \eps )=\int_\eps^\infty~ dz {1\over z} e^{-z} $ 
and $g_S={4\pi^2 G_S/ N_c \Lambda^2}$. 

As an example, at next to leading order in $1/N_c$ the resummation 
of the self-energy 
insertion diagrams as in Fig.\tref{fig2}(a) is given by $g_S \Pi =
{g_S \bar\Pi\over 1 - g_S \bar\Pi} + {1\over 1 - g_S \bar\Pi}g_s 
\Sigma {1\over 1 - g_S \bar\Pi} + \cdots$ and one gets
\beq
\rlabel{Pi} 
\Pi(Q^2)=-{1\over g_S}\left(1 - {(Z_S(Q^2) g_S)^{-1}
\over  Q^2+(2M_Q)^2- \Sigma(Q^2)Z_S(Q^2)^{-1}}\right) ,
\eeq
with
\beq
\rlabel{Sigma} 
\Sigma(q^2)= {1\over 2}g_S^2 \int {d^4k\over (2\pi)^4}
{\bar T(q,k)}^2 {1 \over 1 - g_S \bar \Pi(k^2)}\, {1 \over 1 - g_S \bar
\Pi((q-k)^2)} . 
\eeq
${\bar T(q,k)}$ is a three-point function vertex of the type 
SSS, SPP, SVV, SAA, SPA (S=scalar, P=pseudoscalar, V=vector, 
A=axial-vector).

Because of the appearance of overlapping divergences and the necessity 
of numerically evaluating contributions like \rref{Sigma} due to a 
complex momenta dependence we chose to estimate them using a 
reliable and fully analytical approximation within the bosonized version 
of the model.
The correspondence with the non bosonized case
is such that a string of quark bubbles is replaced by 
a meson line with the same quantum numbers. The n.t.l. $1/N_c$ 
corrections become one loop corrections in the meson theory.
In the bosonized version, after integrating over constituent quarks, 
only the auxiliary boson fields remain: scalar and pseudoscalar in the 
$G_V=0$ case and the additional vector and axial-vector fields
in the $G_V\neq 0$ case. In what follows we refer to the bosonized 
version with the non linear realization of the chiral symmetry
(i.e. the non linear representation for the pseudoscalar 
field and derivative coupling of the pseudoscalar to the other degrees 
of freedom).

Our approximation does correspond in practice to neglecting momenta 
dependence in vertices and masses of \rref{Pi}. 
We discuss in Section \tref{2} the numerical relevance of the 
approximation.
On the more formal side our approximation corresponds to compute the 
next to leading $1/N_c$ corrections within the bosonized 
version keeping the 
leading order contributions 
in the Heat Kernel Expansion (HKE) approach \rcite{ENJL}
to the boson vertices and masses. 
(see also \rcite{BALL} for a review on HKE).
Besides this it is easy to verify that the resummed HKE for a given 
interaction vertex 
(which is an expansion in powers of $\partial^2/M_Q^2$)
and the large-$N_c$ resummation of quark bubbles produce the SAME
momenta dependence.

For the concerns of the numerical evaluation it is useful to notice that
the HKE behaves as a slowly convergent series (alternating 
signs with slowly decreasing coefficients) which implies that the 
leading term is a better estimate of the exact result than any 
truncation at a finite 
order outside the domain $q^2\ll M_Q^2$. In what follows, 
only the lowest order in the derivative expansion will be kept
for each vertex.
This approximation allows us to simplify the calculations and preserves
chiral invariance.

\section{\bf The $G_V=0$ case.}

\setcounter{equation}{0}
\rlabel{2}

In the $G_V=0$ case non renormalizable four-quark interactions in 
the Lagrangian \rref{ENJL} reduce to the scalar and pseudoscalar 
type with one 
coupling constant $G_S$. The bosonization introduces scalar and pseudoscalar 
auxiliary fields and the integration over constituent quarks generates 
the effective action for the scalar and pseudoscalar physical mesons.
The pseudoscalar sector is the ChPt Lagrangian of the pseudo-Goldstone 
bosons.
The effective Lagrangian thus obtained is by 
construction globally chiral invariant  
(and locally chiral invariant in presence of external 
left and right handed sources) and it is non renormalizable being an 
infinite expansion in powers of derivatives acting on the meson fields. 
Details on the method can be found in \rcite{ENJL}.

We restrict ourselves to the $U(2)_L\times U(2)_R$ case (we disregard
the effect of the $U(1)_A$ anomaly which is also a next to leading 
effect in $1/N_c$).
The general form of the meson fields, singlets or triplets under
$SU(2)_V$, reads
\beq M= \sum_{a=1}^3 {1\over \sqrt{2}} M_{(a)}\tau^{(a)} + {1\over 
\sqrt{2}}M_0 {\bf 1}, 
\rlabel{taum}
\eeq
where $\tau^{(a)}, a=1,..3$ are the Pauli matrices with $Tr 
(\tau^a\tau^b) = 2\delta^{ab}$ and $M_0$ is the singlet component.
In the chiral limit ($m_u=m_d=0$), the effective chiral Lagrangian
including scalar and pseudoscalar mesons at leading order $O(p^2)$
in the derivative expansion  is given by:
\ba
{\cal{L}}^{S,P}&=&{f_\pi^2\over 4} <\xi_\mu \xi^\mu>
+{1\over 2}<d_\mu S d^\mu S>
-{1\over 2}M_S^2<S^2>+~{\cal{L}}_{int}^{S,P} \nonumber\\
{\cal{L}}_{int}^{S,P}&=&-{\lambda_3\over 3!}<S^3> - {\lambda_4\over
4!}<S^4>   + c_d <S\xi_\mu\xi^\mu > + c_4^{(1)} <S^2\xi_\mu\xi^\mu >
\nonumber\\
&&+ c_4^{(2)} <S\xi_\mu S\xi^\mu > .
\rlabel{LSP}
\ea
Building blocks of the Lagrangian \rref{LSP} are the scalar field $S$
and the axial current of the pseudoscalar field $\xi_\mu = i\{\xi^\dagger
(\partial_\mu -ir_\mu )\xi - \xi (\partial_\mu -il_\mu )\xi^\dagger\}$,
where $r_\mu$ and $l_\mu$ are the external right-handed and left-handed 
sources
and $\xi =\sqrt{U} = \exp (-{i\over \sqrt{2}}{\Phi\over f_\pi})$ is the 
usual exponential representation with the pseudoscalar meson matrix 
$\Phi$ defined as in \rref{taum}.
Both fields $\xi_\mu$ and $S$ transform non linearly under the chiral 
group $G=U(2)_L\times U(2)_R$ as ${\cal{O}}\to h(\Phi ){\cal{O}} h^\dagger 
(\Phi )$.
The couplings amongst mesons have been derived using the HKE
and with proper time regularization.  
Their expressions are listed in Appendix A.
They are functions of the cut-off $\Lambda$ of the fermion loop,
the constituent quark mass $M_Q$, the axial-pseudoscalar mixing parameter
$g_A$ ($g_A=1$ in the case $G_V=0$) and  the number of colours $N_c$. 

As it is implied by the non renormalizability of the model the values of 
the parameters are {\em a priori} regularization dependent. Most 
suitable regularizations are the covariant ones: proper time, 
four-momentum cut-off and Pauli-Villars. Explicit solutions of the gap 
equation of the four-fermion model in the three cases can be 
found in \rcite{KLEVANSKY}. A small 
regularization dependence of the parameters has been found
\rcite{KLEVANSKY,GRANDE}.

\begin{figure}
\begin{center}
\leavevmode\epsfxsize=14cm\epsfysize=2.8cm\epsfbox{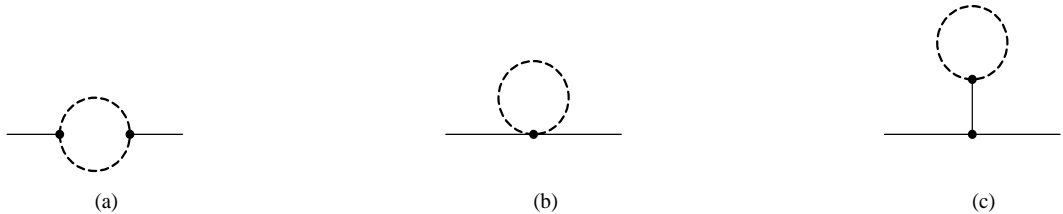}
\end{center}
\caption{Two-point scalar function in the bosonized ENJL model at next to 
leading order in $1/N_c$. Diagrams (a) are self-energy insertions. 
Diagrams (b) are tadpoles. Diagrams (c) contribute to the one-point 
scalar function and have to be included in the gap equation.}
\label{fig3}
\end{figure}

The next to leading in $1/N_c$ corrections to the pole mass 
of the scalar two-point function 
within the bosonized version are given by the one loop 
corrections to the scalar 
meson propagator generated by the vertices in 
\rref{LSP}.
The diagrams which contribute are the ones in Fig.\tref{fig3}(a),
the self-energy insertions and \tref{fig3}(b), the tadpoles. 
The diagram 
\tref{fig3}(c) does enter the gap equation (it modifies the one point 
scalar function) and has not to be included in order to avoid double 
counting. The $1/N_c$ corrections to the gap equation have been 
considered in \rcite{GRANDE} and proven to be numerically relevant but 
still in a perturbative regime. The one boson loops have to be 
regularized and thus explicitly
depend on a {\em new} cut-off $\tilde\Lambda$. This is the signal of the 
non renormalizability of the model. Physical inputs can be 
used to constrain its value (see section \tref{NUM}).
One loop diagrams in Fig.\tref{fig3} can be up to quartically 
divergent by naive power counting
due to the derivative coupling of the pseudoscalar field.
As was noticed in \rcite{Metz} with a specific example of a 
phenomenological pion Lagrangian \`a la Weinberg, the quantization of 
effective theories like the one in \rref{LSP} with an arbitrary number 
of derivatives, can fail if one uses naive Feynman rules with a cutoff 
regularization scheme.

In the next subsection we briefly show that all the leading quartic 
divergences disappear under
the requirement of chiral invariance of the partition 
function. Our case is a simple extension of the 
example shown in \rcite{Metz} to an effective theory of pions interacting 
with scalar fields.

\subsection{Quartic divergences versus chiral invariance}

\rlabel{quartic}

The Lagrangian \rref{LSP} satisfies two requisites: a) it is an infinite 
expansion in powers of derivatives acting on the fundamental fields and
b) the fundamental fields transform non linearly under the chiral group.
Expanding the field $\xi_\mu$ in powers of the $\Phi$ field matrix, 
and reducing covariant derivatives to ordinary ones which enter in our 
calculation, the Lagrangian \rref{LSP} can be written as 
${\cal{L}}={1\over 2} g^{\tilde{a}\tilde{b}}(\Phi ,S)
\partial_\mu\Phi^{\tilde{a}}\partial^\mu\Phi^{\tilde{b}}$, 
where $\tilde{a}, \tilde{b}$ are flavour indices going 
from 0 (for the singlet case) to 3
(for the triplet case) according to the decomposition \rref{taum}.
The metric tensor $g^{\tilde{a}\tilde{b}}(\Phi ,S)$ 
is explicitly dependent on the 
pseudoscalar field and the scalar field due to the presence of
interaction terms. We find

\ba
g^{\tilde{a}\tilde{b}}&=&
\delta^{\tilde{a}\tilde{b}}\biggl [ 1+{4c_d\over \sqrt{2}f_\pi^2} S_1
+{2\over f_\pi^2} (c_4^{(1)}+c_4^{(2)}) S_1^2 
+{2\over f_\pi^2} (c_4^{(1)}-c_4^{(2)}) \sum_{i=1}^3 S_iS^i\biggr ]
\nonumber\\
&&+ (S_a\delta^{0\tilde{b}}+S_b\delta^{0\tilde{a}})
\biggl ( {4c_d\over \sqrt{2}f_\pi^2} +
{4\over f_\pi^2} (c_4^{(1)}+c_4^{(2)}) S_1\biggr )
\nonumber\\
&&+{4\over f_\pi^2} c_4^{(2)} \sum_{i=1}^3 S_iS^i 
\delta^{0\tilde{a}}\delta^{0\tilde{b}}
+ {4\over f_\pi^2} c_4^{(2)}S^aS^b,
\ea 
with flavour indices $a, b$=1,...3.
The metric tensor $g^{\tilde{a}\tilde{b}}$ defines a non linear chiral
 transformation
of the pseudoscalar field $\Phi$ contained in the original Lagrangian 
${\cal{L}}={1\over 2} \delta^{\tilde{a}\tilde{b}}
\partial_\mu\Phi^{\tilde{a}}\partial^\mu\Phi^{\tilde{b}}$
with a flat metric $\delta^{\tilde{a}\tilde{b}}$. 
Under this transformation the full partition 
function has to be invariant (we are not concerned with anomaly in this 
context). 
Since the chiral transformed measure is the original one multiplied by 
$\delta^4(0)\sqrt{det g^{\tilde{a}\tilde{b}}}$, the chiral invariant 
partition function is defined in terms of the new 
Lagrangian ${\cal{L}}^\prime = 
{\cal{L}} + \delta^4(0)\ln \sqrt{det g^{\tilde{a}\tilde{b}}}$. 
By doing an expansion in the small couplings to the scalar field we find

\ba
\ln \sqrt{det g^{\tilde{a}\tilde{b}}} &=&
{8c_d\over \sqrt{2}f_\pi^2} S_1 
- \biggl ({4c_d\over \sqrt{2}f_\pi^2}\biggr )^2 
\biggl ( S_1^2+{1\over 2} \sum_{i=1}^3 S_iS^i\biggr )  
\nonumber\\
&&+{4\over f_\pi^2} c_4^{(1)} \sum_{i=1}^3 S_iS^i 
+ {4\over f_\pi^2} (c_4^{(1)}+c_4^{(2)}) S_1^2 +O(S^3).
\ea
The new terms with $\delta^4(0)=\int~d^4 k/(2\pi)^4$ do exactly cancel
all the leading quartic divergences generated by diagrams in 
Fig.\tref{fig3} 
whose final expressions are listed in Appendix B.
The first term cancels the quartic divergence of Fig.\tref{fig3}(c) 
(referred to as ``top'' diagram in Appendix B) for the 
pseudoscalar case, the second term cancels the one in Fig.\tref{fig3}(a) 
(self-energy) pseudoscalar,
and the last two terms the one in Fig.\tref{fig3}(b) (tadpole) pseudoscalar
in the triplet and singlet case respectively.

\subsection{ Numerical analysis}

\rlabel{NUM}

We have calculated the $1/N_c$ corrections in two cases: 
1) assuming that the
scalar particle is a singlet 
 and 2) assuming that the quark
content of the scalar particle is the same as that of the $\rho (770)$
vector meson. This could be the case of the physical $a_0 (983)$ scalar
resonance. Obviously our $SU(2)$ calculation has to be
interpreted as  a first indicative approximation of the fully realistic
$SU(3)$  calculation.
The self-energy and tadpole contributions are listed in Appendix B for 
the scalar and pseudoscalar loops and both for the singlet and triplet
cases.
In the chiral limit ($m_\pi =0$) all the pseudoscalar one loop corrections 
vanish. Denoting with $M_S^2=(2M_Q)^2$ the pole mass of the scalar 
two-point function in the large-$N_c$ limit, the corrected scalar 
mass at next-to-leading order in $1/N_c$ with a proper time 
regularization is the following in the singlet case:
\beq
\tilde{M}^2_{S_1} = M_S^2\biggl [1+{\lambda_4\over 16\pi^2}
\Gamma (-1,M_S) -{\lambda_3^2\over 16\pi^2}{1\over 
M_S^2} \Gamma (0,M_S)\biggr ] ,
\rlabel{CONT1}
\eeq
while for the neutral scalar triplet (the one associated with $\tau_3$)
we get
\beq
\tilde{M}^2_{S_3} = M_S^2\biggl [ 1 + {2\over 3}{\lambda_4\over 16\pi^2}
\Gamma (-1,M_S)-{1\over 2}{\lambda_3^2\over 16\pi^2}{1\over 
M_S^2} \Gamma (0,M_S)\biggr ]  .
\rlabel{CONT2}
\eeq
Away from the chiral limit (i.e. $m_\pi\neq 0$)
the additional corrections we get are as follows:
\ba
\Delta\tilde{M}^2_{S_1}&=& {M_S^2\over 16\pi^2}{2m_\pi^2\over f_\pi^2}
\biggl [ -{4c_d^2\over f_\pi^2}\biggl ( 1-4{m_\pi^2\over M_S^2}\biggr )
\Gamma (-1,m_\pi )
-4(c_4^{(1)}+c_4^{(2)}){m_\pi^2\over M_S^2}\Gamma (-1,m_\pi ) 
\nonumber\\
&&+{8c_d^2\over f_\pi^2}\biggl ( 1-{m_\pi^2\over M_S^2}\biggr )
\Gamma (0,m_\pi )\biggr ] ,
\rlabel{CONT3}
\ea
\ba
\Delta\tilde{M}^2_{S_3}&=& {M_S^2\over 16\pi^2}{2m_\pi^2\over f_\pi^2}
\biggl [ -{2c_d^2\over f_\pi^2}\biggl ( 1-4{m_\pi^2\over M_S^2}\biggr )
\Gamma (-1,m_\pi )
-4c_4^{(1)}{m_\pi^2\over M_S^2}\Gamma (-1,m_\pi )\hspace{.6cm} 
\nonumber\\
&&+{4c_d^2\over f_\pi^2}\biggl ( 1-{m_\pi^2\over M_S^2}\biggr )
\Gamma (0,m_\pi )\biggr ] ,
\rlabel{CONT4}
\ea
which also include the contribution from the wave function 
renormalization constant.
We notice however that explicit mass terms in the pseudoscalar 
Lagrangian have not been included.
The partial gamma functions $\Gamma (n,M_i)$ of the proper time 
regularization depend upon the adimensional ratio 
$M_i^2/\tilde{\Lambda}^2$, with $i=S,P$ and where $\tilde{\Lambda}$ is the 
{\em new} cut-off of the one-boson loop.
It is worth to notice that the diagrams in Fig.\tref{fig3}(c) (top)
which have not been included here do not generate any 
mass splitting between the singlet and the triplet scalar component
as expected for a contribution to the gap equation, 
while the self-energy diagrams give to the triplet component half of the 
contribution to the singlet one.

We have disregarded  the splitting between the singlet and the triplet 
components running in the loops. In the pseudoscalar case this is due
to the $U(1)$ axial anomaly, which appears in the effective Lagrangian 
at next-to-leading order in the $1/N_c$ expansion. In the scalar case it 
is again a next-to-leading effect in $1/N_c$ as we have shown here, 
although other sources can 
compete in this sector like mixing with glueballs.

The numerical evaluation of the $1/N_c$ corrections in 
\rref{CONT1}, \rref{CONT2}, \rref{CONT3} and \rref{CONT4} 
needs as input the values of the large-$N_c$ parameters 
of the ENJL model, $\Lambda$ and $M_Q$ (or alternatively $\Lambda$ and 
$G_S$) and the {\em new} one-loop cut-off $\tilde{\Lambda}$. All 
these quantities are regularization dependent and have to be 
consistently evaluated
in the same scheme. We used the proper time regularization,
while the corresponding expressions in the Pauli-Villars scheme 
can be easily 
obtained (previous cancellation of the spurious leading divergences 
due to the non invariance of the measure)
through the substitutions listed at the end of Appendix B.
For the choice of the numerical value of $\tilde{\Lambda}$ we 
follow the argument 
developed in \rcite{GRANDE} which, although purely phenomenological, 
provides a self-consistent way of estimating the size of the boson loop 
cut-off; it proves that keeping the 
physical value of $f_\pi$ at n.t.l. order in $1/N_c$
constrains the allowed range for $\tilde{\Lambda}$ to be 
$\tilde{\Lambda}\leq \tilde{\Lambda}_{max}$, where $\tilde{\Lambda}_{max}$
is of the order of the constituent quark loop cut-off $\Lambda$.
The values for the large-$N_c$ parameters in the proper time 
regularization are $M_Q=199$ MeV and $\Lambda = 667$ MeV in the $G_V=0$ 
case (see fit 4 of \rcite{BIJNENS,ENJL}). 
The analysis in \rcite{KLEVANSKY} shows in addition
a small dependence of these parameters upon the regularization scheme.
\begin{figure}
\rotate[l]{\epsfxsize=7cm\epsfysize=14cm\epsfbox{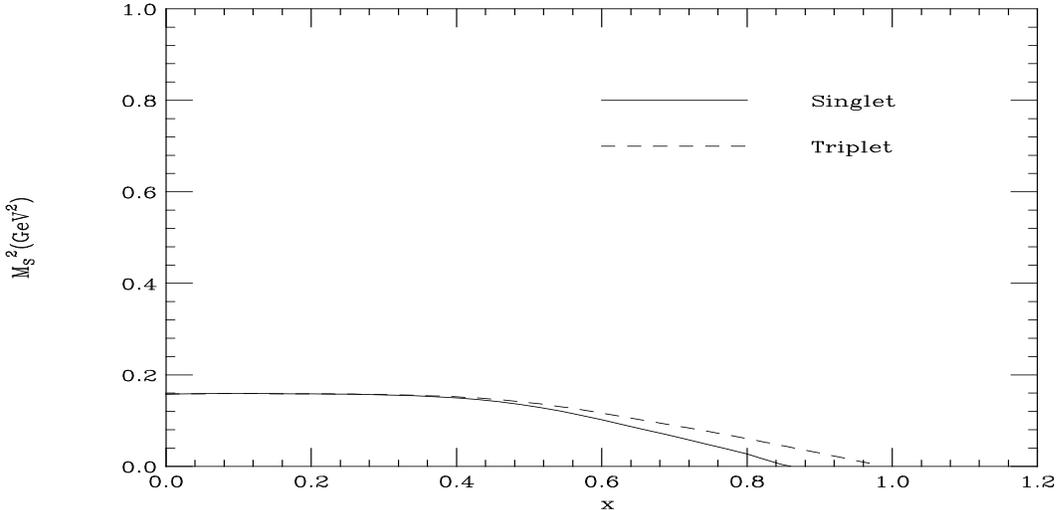}}
\caption{ The squared scalar mass at next-to-leading order in $1/N_c$
in the case $G_V=0$ as a function of the ratio 
$x=\tilde\Lambda/\Lambda$ for the singlet case (solid curve)
and the triplet case (dashed curve). Here $\Lambda$ is fixed at
$\Lambda =667$ MeV and $M_Q=199$ MeV according to fit 4 of 
\protect{\rcite{BIJNENS}}}
\label{figpl1}
\end{figure}
In Fig.\tref{figpl1} we show the squared scalar mass 
corrected at next to leading order in $1/N_c$ and in the chiral limit
(formulas \rref{CONT1} and \rref{CONT2})
in the singlet and triplet cases as a function of the boson loop cut-off
$\tilde{\Lambda}$ with fixed $M_Q=199$ MeV, $\Lambda =667$ MeV. 
The scalar boson mass $M_S^2$ in the r.h.s of \rref{CONT1} and 
\rref{CONT2} is fixed at its large-$N_c$ value $M_S^2=4 M_Q^2$.  
The corrections are negative both to the singlet and the triplet
states
and push the mass to zero already at $\tilde\Lambda/\Lambda \simeq 0.8$.
A triplet-singlet splitting is induced which grows with $\tilde\Lambda$ 
but remains small.
Away from the chiral limit, with the physical pion mass, pseudoscalar
contributions are again negative but suppressed.
The behaviour of the genuine next to leading $1/N_c$ corrections 
seems to be in qualitative agreement with the results based on 
equivalence arguments as in \rcite{ZJ,Akama}.
\begin{figure}
\rotate[l]{\epsfxsize=7cm\epsfysize=14cm\epsfbox{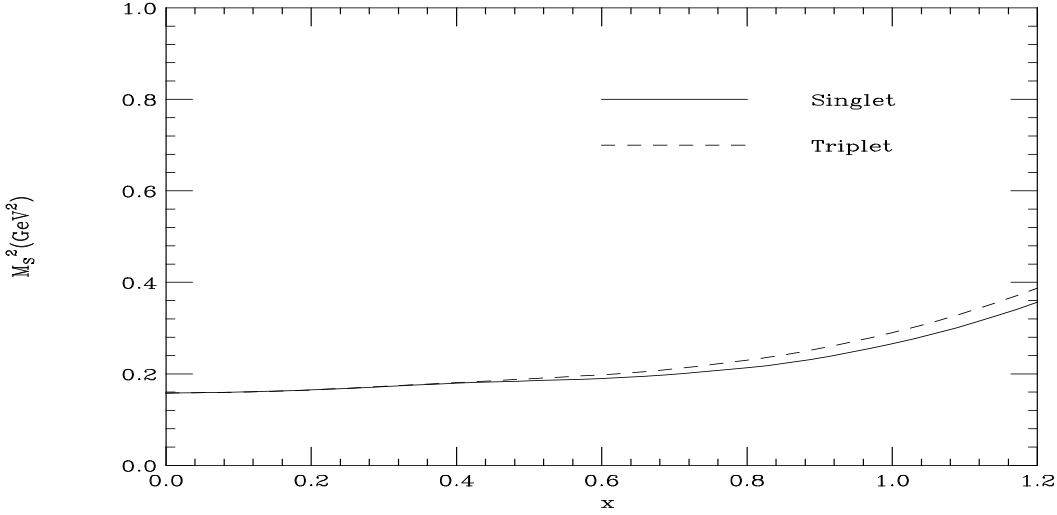}}
\caption{The partially resummed scalar mass squared 
using $M_Q(\tilde\Lambda )$
corrected at next-to-leading order in $1/N_c$
in the case $G_V=0$ as a function of the ratio 
$x=\tilde\Lambda/\Lambda$ for the singlet case (solid curve)
and the triplet case (dashed curve). Here $\Lambda$ is fixed at
$\Lambda =667$ MeV and $M_Q(x)=0.199+.0995x^2$ GeV which reproduces 
a $50\%$ of 
positive correction to its leading $N_c$ value at $x=1$
according to the results in \protect{\rcite{GRANDE}} }
\label{figpl2}
\end{figure}
The interesting exercise is to take into account the n.t.l. 
$1/N_c$ corrections
to the constituent quark mass $M_Q$ as a solution of the gap equation.
This induces a partial resummation of the $1/N_c$ corrections to the 
scalar boson mass. 
The $1/N_c$ corrections to the gap equation have been already computed 
in \rcite{GRANDE} and they cause a positive shift
of the constituent quark mass $M_Q$ as a function of $\tilde\Lambda$ for 
fixed $\Lambda$ and $f_\pi$.
In Fig.\tref{figpl2} we show 
the result of using a running value of $M_Q(\tilde\Lambda )$ in formulas
\rref{CONT1} and \rref{CONT2}) which qualitatively reproduces the 
behaviour found in \rcite{GRANDE}.
The surprising result is that the partially resummed corrections are now 
positive and softer, while the splitting is not modified.

To estimate the error which affects our zero momenta approximation
we studied the momentum dependence of each vertex entering the boson 
loop.
All the couplings are weakened by the $q^2$ corrections and reduced in 
absolute value by 
about $20\div 30\%$ up to $-q^2\simeq\Lambda^2$.
This leads to the conclusion that the $q^2$ resummed value 
cannot overcome the approximated value.
The same numerical results for the scalar mass in the Pauli-Villars 
regularization
are obtained to a good approximation with the rescaling of the proper-time 
cut-off $\tilde\Lambda_{PT}\simeq 2\sqrt{2}\tilde\Lambda_{PV}$.

\section {The $G_V\neq 0$ case.}

\setcounter{equation}{0}
\rlabel{3}

The less explored behaviour of four-fermion models is in the presence
of vector like interactions, i.e. $G_V\neq 0$ in our case.
The Interaction Lagrangian of
scalar mesons with vectors and  axial-vectors at leading order in the 
derivative expansion is:
\begin{eqnarray} 
{\cal{L}}_{int}^{V,A}&=&
c_V^{(1)}<SV_\mu SV^{\mu} >+c_V^{(2)} <S^2V_\mu V^{\mu}>
+ c_{AP} <S\{\xi_\mu ,A^\mu\} >\nonumber\\ 
&&+\tilde{c}_A <SA_\mu A^{\mu} > 
+ c_A^{(1)}<SA_\mu SA^{\mu} >+ c_A^{(2)} <S^2A_\mu A^{\mu} >.
\end{eqnarray}
All the couplings are listed in Appendix A.
Notice also the presence at $O(p)$ of the mixed term 
scalar-pseudoscalar-axial with coupling $c_{AP}$. 
The additional diagrams contributing to the scalar pole mass 
are again the ones in Fig.\tref{fig3}(a), (b) 
with vector, axial, or mixed axial-pseudoscalar
internal lines. All the one loop contributions are 
listed in Appendix B. In this case quartic divergences can be addressed 
to two different sources: a) for diagrams with derivative couplings 
their origin can be the breaking of chiral invariance as for the 
genuine pseudoscalar case, b) for diagrams with non derivative couplings 
quartic divergences are a natural consequence of the 
the bad high energy behaviour of the massive vector propagator 
$\Delta_{\mu\nu} = (g_{\mu\nu}-k_\mu k_\nu /M_V^2) 
/(k^2-M_V^2)$ and they signal the non 
renormalizability of the massive vector Lagrangian.
\begin{figure}
\rotate[l]{\epsfxsize=7cm\epsfysize=14cm\epsfbox{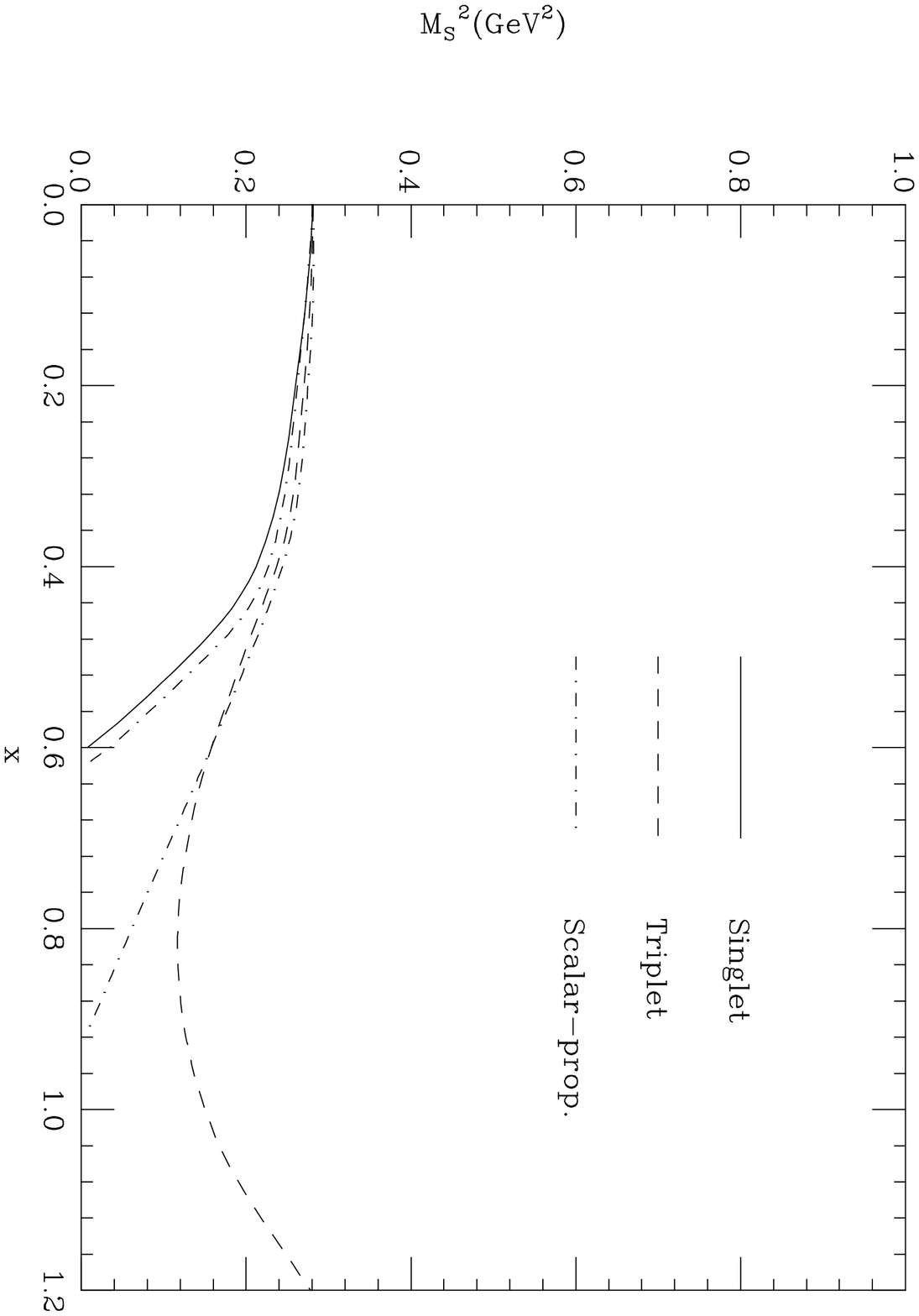}}
\caption{The squared scalar mass at next-to-leading order in $1/N_c$
in the case $G_V\neq 0$ as a function of the ratio 
$x=\tilde\Lambda/\Lambda$ for the singlet case (solid curve) and
the triplet case (dashed curve). The effect of using the 
scalar-like vector (axial) propagator is also shown
(dot dashed curve). 
Here $\Lambda$ is fixed at
$\Lambda =1.16$ GeV, $g_A=0.61$ and $M_Q=265$ MeV according 
to fit 1 of \protect{\rcite{BIJNENS}} }
\label{figpl3}
\end{figure}
Divergences of type a) are cancelled following the same demonstration 
as in \tref{quartic} where the generic field $\Phi$ is now replaced 
by a generic vector field $V_\mu$. They are absent in our case.
Divergences of type b) can be cured by the introduction of a spontaneous 
symmetry breaking mechanism or taking into account the compositness 
of the vector fields. Nonetheless we observe that a nearly quartic 
divergence could not be avoided in the calculation within the non 
bosonized version using the large $N_c$ vector two-point functions
predicted in \rcite{2point}, where the running vector mass behaves like 
$M_V(k)\sim \ln k$. This is the signal of the expected bad high energy
behaviour of an effective NJL model.
In Appendix B we show the results obtained 
using the ordinary propagator of a massive
vector field $\Delta_{\mu\nu} = (g_{\mu\nu}-k_\mu k_\nu /M_V^2) 
/(k^2-M_V^2)$. 
As an example we also studied the results for the scalar-like
propagator with 
softer renormalizable high energy behaviour 
$\Delta_{\mu\nu} = g_{\mu\nu}/(k^2-M_V^2)$. 
In the chiral limit and for the non renormalizable massive vector 
(axial)
propagator the additional corrections to the scalar mass are as follows:
\ba
&&\Delta\tilde{M}^2_{S_1}=\nonumber\\ 
&&{4M_S^2\over 16\pi^2}\biggl\{
-{\tilde{c}_A^2\over M_S^2}\biggl [2\Gamma (-2, M_A)+
\biggl ( 1+{M_S^2\over 2M_A^2}\biggr )\Gamma (-1, M_A)
+\biggl ( 3-{M_S^2\over 2M_A^2}\biggr )\Gamma (0, M_A)\biggr ]
\nonumber\\
&&-2{c_{AP}^2\over f_\pi^2}{M_A^2\over M_S^2}
\biggl [\Gamma (-2, M_A)+
\biggl ( 1-{M_S^2\over 2M_A^2}\biggr )\Gamma (-1, M_A)
+\biggl ( {1\over 2}-{M_S^2\over M_A^2}\biggr )
\nonumber\\
&&\int_0^1~d\alpha~\Gamma \biggl ( 0, {\alpha M_A^2\over 
\tilde{\Lambda}^2}\biggr )\biggr ]
+ (c_A^{(1)}+c_A^{(2)}){M_A^2\over M_S^2} \biggl [2\Gamma (-2, M_A)+4
\Gamma (-1, M_A)\biggr ]\biggr\} ,
\rlabel{CONTV3}
\ea

\begin{figure}
\rotate[l]{\epsfxsize=7cm\epsfysize=14cm\epsfbox{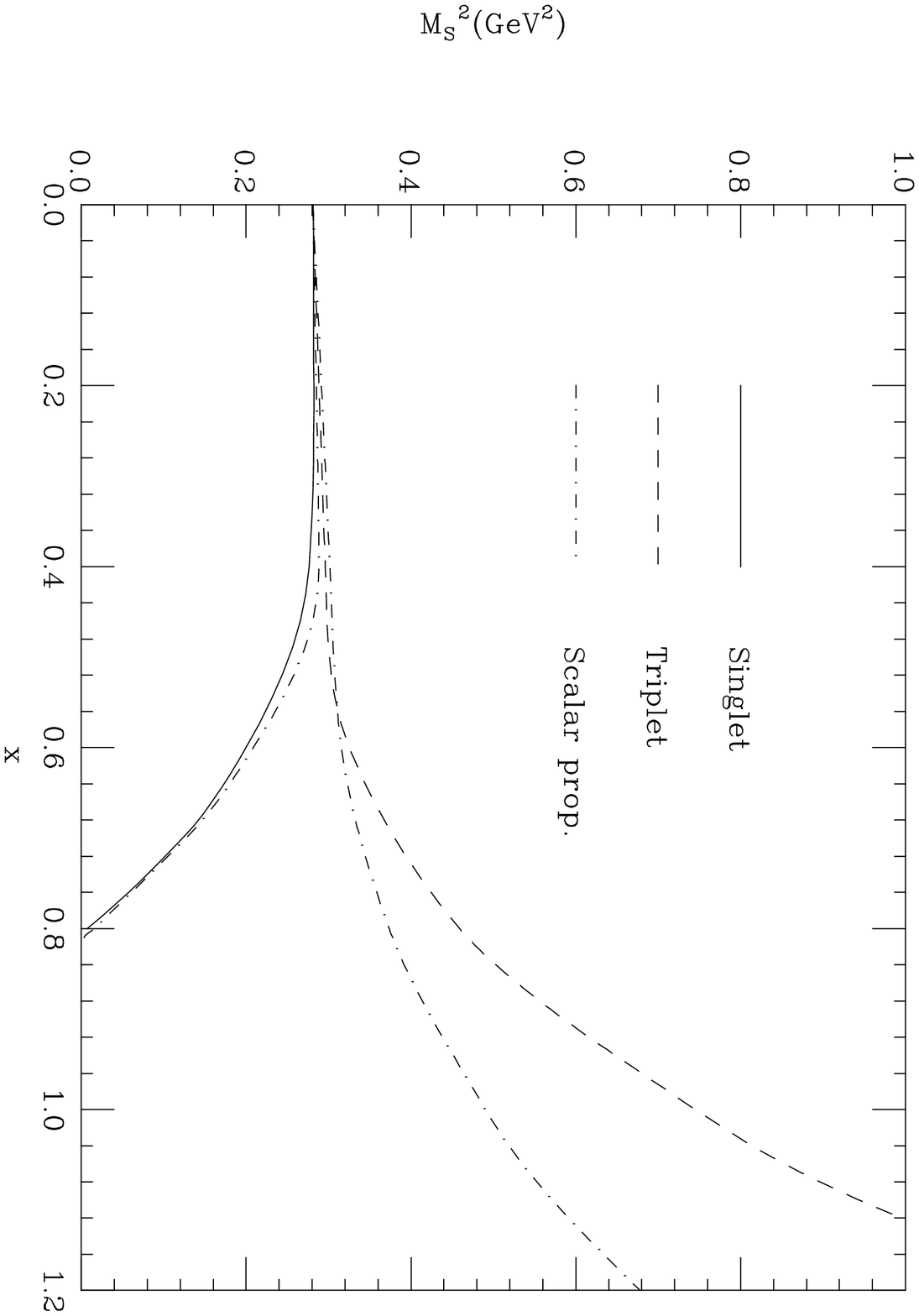}}
\caption{The squared scalar mass using $M_Q(\tilde\Lambda )$
corrected at next-to-leading order in $1/N_c$
in the case $G_V\neq 0$ as a function of the ratio 
$x=\tilde\Lambda/\Lambda$ for the singlet case (solid curve)
and the triplet case (dashed curve). The effect of using 
the scalar-like vector (axial) propagator is also shown
(dot dashed curve). Here $\Lambda$ is fixed at
$\Lambda =1.16$ GeV $g_A=0.61$ 
and $M_Q(x)=0.265+.1325x^2$ GeV which reproduces 
a $50\%$ of 
positive correction to its leading $N_c$ value at $x=1$
according to the results in \protect{\rcite{GRANDE}} }
\label{figpl4}
\end{figure}

\ba
&&\Delta\tilde{M}^2_{S_3}=\nonumber\\
&&{4M_S^2\over 16\pi^2}\biggl\{
-{1\over 2}{\tilde{c}_A^2\over M_S^2}\biggl [2\Gamma (-2, M_A)+
\biggl ( 1+{M_S^2\over 2M_A^2}\biggr )\Gamma (-1, M_A)
+\biggl ( 3-{M_S^2\over 2M_A^2}\biggr )\Gamma (0, M_A)\biggr ]\nonumber\\
&&-{c_{AP}^2\over f_\pi^2}{M_A^2\over M_S^2}
\biggl [\Gamma (-2, M_A)+
\biggl ( 1-{M_S^2\over 2M_A^2}\biggr )\Gamma (-1, M_A)
+\biggl ( {1\over 2}-{M_S^2\over M_A^2}\biggr ) \cdot\nonumber\\
&&\int_0^1~d\alpha~\Gamma \biggl ( 0, {\alpha M_A^2\over 
\tilde{\Lambda}^2}\biggr )\biggr ] 
+ c_A^{(2)}{M_A^2\over M_S^2} \biggl [2\Gamma (-2, M_A)+4
\Gamma (-1, M_A)\biggr ]
\nonumber\\
&&+ c_V^{(2)}{M_V^2\over M_S^2} \biggl [2\Gamma (-2, M_V)+4
\Gamma (-1, M_V)\biggr ]
\biggr\} .
\rlabel{CONTV4}
\ea
As in the $G_V=0$ case,
we studied the pure next to leading $1/N_c$ corrected scalar mass
and the partially resummed one using for the vector and axial masses
$M_V=0.8$ GeV and $M_A=1$ GeV.
The first result is shown in Fig.\tref{figpl3}. 
The large-$N_c$ values of the  parameters
in the $G_V\neq 0$ case with proper time regularization
are $M_Q=265$ MeV, $\Lambda =1.16$ GeV and $g_A=0.61$ (see fit 1 in 
 \rcite{ENJL,BIJNENS}).
A few comments are in order. The singlet-triplet splitting is enhanced 
respect to the $G_V=0$ case. The singlet mass still receives negative 
corrections. The anomalous enhancement of the triplet mass is sensitively 
dependent on the form of the propagator. It is consequence 
of the presence of positive contributions in the vector sector which 
actually dominate in the case of the ordinary vector propagator form and 
that are zero in the singlet case.
The partially resummed behaviour is shown in Fig.\tref{figpl4}. 
Corrections to the singlet state are softened but still negative.
The same anomalous enhancement of the triplet mass is observed.
In both cases the singlet-triplet splitting is enhanced respect to the 
$G_V=0$ case.
Again, on the base of the study of the $q^2$ dependence of the vector 
(axial) vertices we expect that the inclusion of the full $q^2$ 
dependence will soften the corrections.
Within the present approximation the largeness of the axial and vector 
corrections prevents from a fully reliable estimate in the region 
$\tilde\Lambda/\Lambda\simeq 1$.

\section {\bf Conclusions} 
\rlabel{CONC}

We studied the next-to-leading in $1/N_c$ corrections to the pole mass of 
the scalar two-point function within the bosonized version of the 
Extended NJL model and away from the infrared domain. 
In this context the model is treated as fully non renormalizable and a new 
cut-off parameter have to be introduced for the one boson loop.
Within a reliable zero momenta approximation, which is the leading 
order of the Heat Kernel Expansion, we have analytically derived
the next-to-leading $1/N_c$ corrections to the scalar mass in both the 
$G_V=0$ and $G_V\neq 0$ (vector and axial fields present) cases and 
studied their regularization scheme dependence.
The main results are that genuine next to leading $1/N_c$ corrections 
to the singlet state 
are negative and relatively large, while a partially resummed estimate
induces positive and softened corrections in the $G_V=0$ case.
Remarkably the corrections to the large-$N_c$ degenerate mass for the 
triplet and singlet states induce a splitting which mimics the physical
one (octet heavier than the singlet).
The splitting effect is enhanced in the $G_V\neq 0$ case.

The largeness of the negative pure next to leading $1/N_c$
corrections derived in this framework 
qualitatively agrees with the results for the scalar over fermion mass 
ratio derived in the IR limit where the equivalence with renormalizable 
Yukawa-type models is valid by the use of the 
compositness condition \rcite{Akama}. 
This suggests an asymptotic behaviour of the $1/N_c$ expansion
of the mass ratios in four-fermion models both at
the IR limit and away from the IR limit, where a 
truncation at any finite order
fails to be a good estimate of the real value for useful values of $N_c$.

\vspace{3.cm} 
{\bf{Acknowledgements}} \vspace{0.8cm}

I thank Christophe Bruno, who contributed to the early stage of this 
work. I thank also K. Akama for a kind and useful correspondence 
on the subject, J. Bijnens for having called my attention 
to this problem and for many useful discussions, P. Hasenfratz for a 
stimulating discussion and E. de Rafael for reading the manuscript.
The work is supported by the EU Contract Nr. ERBCHBGCT 930442.
\vspace{2cm}

\appendix

\def\theequation{\Alph{section}.\arabic{equation}}
\section{The couplings of the bosonized Lagrangian}
\setcounter{equation}{0}

The scalar-pseudoscalar couplings are:
\begin{eqnarray} 
{\lambda_3\over 3!} &=&{N_c\over 16\pi^2}4{M_Q\over
Z_S^{3/2}} \biggl [\Gamma (0,\eps ) -{2\over 3}\Gamma (1,\eps )\biggr ]
\nonumber\\ 
{\lambda_4\over 4!} &=&{N_c\over 16\pi^2}{1\over Z_S^2}
\biggl [\Gamma (0,\eps )-4\Gamma (1,\eps ) +{4\over 3} \Gamma (2,\eps
)\biggr ] \nonumber\\ 
c_d&=&  {N_c\over 16\pi^2} M_Q{2g_A^2\over
\sqrt{Z_S}} \biggl [\Gamma (0,\eps )-\Gamma (1,\eps ) \biggr ]
\nonumber\\ 
c_4^{(1)}&=& {1\over 2} {N_c\over 16\pi^2} {g_A^2\over Z_S}
\biggl [\Gamma (0,\eps ) -{20\over 3}  \Gamma (1,\eps ) +{8\over 3}
\Gamma (2,\eps )\biggr ] \nonumber\\ 
c_4^{(2)}&=&{1\over 2} {N_c\over
16\pi^2} {g_A^2\over Z_S} \biggl [\Gamma (0,\eps ) -{10\over 3}  \Gamma
(1,\eps ) +{4\over 3} \Gamma (2,\eps )\biggr ] . 
\rlabel{COUP}
\end{eqnarray}
The scalar-vector and scalar-axial couplings are:
\begin{eqnarray} 
\tilde{c}_A&=& {1\over 2} {N_c\over 16\pi^2}
{M_Q\over  \sqrt{Z_S}Z_V}
 \biggl [4\Gamma (0,\eps ) -4\Gamma (1,\eps ) \biggr ] \nonumber\\
c_A^{(1)}&=&{1\over 2} {N_c\over 16\pi^2}{1\over Z_S Z_V}
 \biggl [\Gamma (0,\eps ) -{10\over 3} \Gamma (1,\eps ) +{4\over 3}
\Gamma (2,\eps )\biggr ] \nonumber\\  
c_A^{(2)}&=&{1\over 2} {N_c\over
16\pi^2}{1\over Z_S Z_V}
 \biggl [\Gamma (0,\eps ) -{20\over 3} \Gamma (1,\eps ) +{8\over 3}
\Gamma (2,\eps )\biggr ] \nonumber\\  
c_{AP}&=&{1\over 2} {N_c\over 16\pi^2}{g_A M_Q\over \sqrt{Z_S}\sqrt{ 
Z_V}}
 \biggl [ -4\Gamma (0,\eps ) +4 \Gamma (1,\eps ) \biggr ] 
\nonumber\\  
c_V^{(1)}&=&-c_V^{(2)} ={1\over 2} {N_c\over
16\pi^2}{1\over Z_S Z_V}  \biggl [-\Gamma (0,\eps )+{2\over 3} \Gamma
(1,\eps ) \biggr ] .
\end{eqnarray}
All the couplings have been derived within the Heat Kernel Expansion 
with proper time regularization.
$Z_V$ and $Z_A$ are the wave function renormalization constants
of the vector and axial-vector fields
 $Z_V = Z_A = {N_c/ 48\pi^2}\Gamma (0,\eps )$.
The partial gamma functions $\Gamma (n-2, \eps )$, with 
$\eps = M_Q^2/\Lambda^2$, are defined as
$ \Gamma (n-2, \eps )=\int_\eps^\infty~ dz {1/ z} e^{-z} z^{n-2}$.
$\Gamma (-2, \eps )$ contains a quartic divergence, $\Gamma
(-1, \eps )$ a quadratic one, $\Gamma (0, \eps )$ is logarithmically
divergent,  while $\Gamma (n, \eps )$ with $n>0$ are finite.

\section{The one loop contributions}
\setcounter{equation}{0}

Here the contributions of the three classes of diagrams of 
Fig.\tref{fig3} are listed for the scalar (S), pseudoscalar (P), vector (V),
axial (A) or mixed axial-pseudoscalar (A-P) bosons running in the loop.
They are the self-energy diagrams of Fig.\tref{fig3}(a), the tadpole 
diagrams of Fig.\tref{fig3}(b) and the top diagrams of 
Fig.\tref{fig3}(c). In the case of axial and vector loops we give the 
result for the form of the propagator  
$\Delta_{\mu\nu} = (g_{\mu\nu}-k_\mu k_\nu /M_V^2) 
/(k^2-M_V^2)$
and the softer one $\Delta_{\mu\nu}^s = g_{\mu\nu}/(k^2-M_V^2)$.
The self-energy contributions are written in the form $A+Bq^2$
with $q^2$ Minkowskian and where $B$ gives the wave function 
renormalization constant which enters the correction to the scalar mass.

\vskip 1cm

{\large\bf Self-energy diagrams }

\vskip 0.4cm

{\bf Singlet propagator}

\begin{eqnarray} 
S &=& i{\lambda_3^2\over 16\pi ^2}
\Gamma ( 0, M_S) \nonumber\\
P &=& i{4c_d^2\over 16\pi ^2} \biggl ({2\over f_\pi^2}\biggr )^2
 \biggl [
-2m_\pi^4 \Gamma( -1,m_\pi )+m_\pi^4 \Gamma ( 0,m_\pi)\nonumber\\ 
&&+q^2\biggl ( {1\over 2}m_\pi^2\Gamma( -1,m_\pi )-m_\pi^2\Gamma ( 0,m_\pi)
\biggr )\biggr]
\nonumber\\
A\,({\Delta_{\mu\nu}})&=& i{4\tilde{c}_A^2\over 16\pi ^2}
 \biggl
[2\Gamma ( -2,M_A) + \Gamma ( -1, M_A) +3\Gamma  (0, M_A)
\nonumber\\ 
&&+{q^2\over 2M_A^2}\biggl (\Gamma ( -1, M_A)-\Gamma ( 0, M_A)\biggr )
\biggr ] \nonumber\\
A\,({\Delta_{\mu\nu}^s})&=& i{4\tilde{c}_A^2\over 16\pi ^2} \biggl
[  4\Gamma (0, M_A)           \biggr ]
\nonumber\\
A-P\,(\Delta_{\mu\nu})&=& 
i {8c_{AP}^2\over 16\pi^2}{2\over f_\pi^2}
\biggl [ {1\over 2}M_A^2 \Gamma ( -2,M_A)
+{3\over 2}{m_\pi^4\over M_A^2}\Gamma ( -2,m_\pi )
\nonumber\\
&&+M_A^2\biggl ({1\over 2}-{m_\pi^2\over 4M_A^2}\biggr )\Gamma ( -1,M_A)
-{1\over 4}m_\pi^2\Gamma ( -1,m_\pi )
\nonumber\\
&&-\biggl (m_\pi^2-{(m_\pi^2+M_A^2)^2\over 4M_A^2}\biggr 
)\int_0^1~d\alpha~ \Gamma \biggl ( 0, {(1-\alpha )m_\pi^2+\alpha 
M_A^2\over \tilde{\Lambda}^2}\biggr )
\nonumber\\
&&+q^2\biggl ( -{1\over 4}\Gamma ( -1,M_A) +{m_\pi^2\over 4M_A^2}
\Gamma ( -1,m_\pi )
\nonumber\\
&& - {m_\pi^2+M_A^2\over 2M_A^2}
\int_0^1~d\alpha~ \Gamma \biggl ( 0, {(1-\alpha )m_\pi^2+\alpha 
M_A^2\over \tilde{\Lambda}^2}\biggr )  \biggr )\biggr ]
\nonumber\\
A-P\,(\Delta_{\mu\nu}^s)&=&  i{8c_{AP}^2\over 16\pi^2}{2\over f_\pi^2}
\biggl [ M_A^2\Gamma ( -1,M_A) - m_\pi^2 
\int_0^1~d\alpha~ \Gamma \biggl ( 0, {(1-\alpha )m_\pi^2+\alpha 
M_A^2\over \tilde{\Lambda}^2}\biggr ) \biggr ] .
\nonumber\\&&
\end{eqnarray}
Notice that no vertex $SVV$ is allowed.
\vskip 0.4cm

{\bf Triplet propagator}

\vskip 0.4cm

\noindent  The only possible self-energy diagrams for the triplet
propagator contain two different internal lines, one singlet and one
triplet. The
contribution is half the contribution to the singlet propagator displayed
above.

\vskip 1cm

{\large\bf Tadpole diagrams}

\vskip 0.4cm

{\bf Singlet propagator}

\begin{eqnarray} 
S &=&-i{\lambda_4\over 16\pi^2} M_S^2 \Gamma ( -1, M_S),
\nonumber\\
P &=& i{4(c_4^{(1)}+c_4^{(2)})\over 16\pi ^2} {2\over f_\pi^2}
 m_\pi^4\Gamma ( -1, m_\pi) \nonumber\\
A(V)\,(\Delta_{\mu\nu})
&=& -i{4(c_{A(V)}^{(1)}+c_{A(V)}^{(2)})\over 16\pi^2}  M_{A(V)}^2
\biggl [ 2\Gamma (-2,M_{A(V)}) +4 \Gamma (-1,M_{A(V)})\biggr ]
\nonumber\\
A(V)\,(\Delta_{\mu\nu}^s) 
&=& -i{4(c_{A(V)}^{(1)}+c_{A(V)}^{(2)})\over 16\pi^2} M_{A(V)}^2
\biggl [ 4 \Gamma (-1,M_{A(V)})\biggr ] .
\end{eqnarray}
Notice that in the singlet case there are no contributions  from
vector vertices of order $O(p^0)$ like $<S^2V_\mu V^{\mu}>$, because 
$c_V^{(1)}+c_V^{(2)} =0$. 

\vskip 0.4cm

{\bf Triplet propagator}

\vskip 0.4cm

\begin{eqnarray} S &=&-i{2\over 3}{\lambda_4\over 16\pi^2}M_S^2 \Gamma
( -1, M_S) \nonumber\\ 
P &=&i{ 4 c_4^{(1)}\over 16\pi^2}{2\over f_\pi^2}m_\pi^4\Gamma ( -1,
m_\pi) \nonumber\\
A(V) \,(\Delta_{\mu\nu})
 &=&-i{4c_{A(V)}^{(2)} \over 16\pi^2}M_{A(V)}^2  
\biggl [ 2 \Gamma (-2,M_{A(V)})+4  \Gamma
(-1,M_{A(V)})\biggr ] 
\nonumber\\ 
A(V) \,(\Delta_{\mu\nu}^s) 
 &=&-i{4c_{A(V)}^{(2)} \over 16\pi^2}M_{A(V)}^2  
\biggl [ 4 \Gamma
(-1,M_{A(V)})\biggr ] 
\end{eqnarray}

{\large\bf Top diagrams}

\begin{eqnarray} 
S &=&i{\lambda_3^2\over 16\pi ^2}\Gamma ( -1, M_S)\nonumber\\
P &=& -i{2\lambda_3 c_d\over  16\pi^2}{2\over f_\pi^2} {1\over M_S^2} 
 m_\pi^4\Gamma ( -1, m_\pi)\nonumber\\
A\,(\Delta_{\mu\nu})&=& i{2\lambda_3 \tilde{c}_A\over 16\pi^2}
 {M_A^2\over M_S^2} \biggl [2 \Gamma(-2,M_A)+ 4 \Gamma
(-1,M_A)\biggr ] \nonumber\\
A\,(\Delta_{\mu\nu}^s) &=& i{2\lambda_3 \tilde{c}_A\over 16\pi^2} 
{M_A^2\over M_S^2} \biggl [4 \Gamma (-1,M_A)\biggr ] .
\end{eqnarray}

\vskip 0.4cm

\noindent Contributions are the same for singlet and triplet propagator.

\vspace{1cm}

{\bf Pauli Villars Regularization}

\vskip 0.4cm

For the comparison with the proper time regularization contributions
the following substitutions can be performed:
\ba
&&m^2\Gamma (-1,m)\to m^2 [(1+2x)\ln (1+2x)-2(1+x)\ln (1+x)]\nonumber\\
&&\Gamma (0,m)\to 2\ln (1+x)-\ln (1+2x),
\ea
where $x=\tilde\Lambda^2/m^2$. This corresponds to the usual Pauli 
Villars procedure in a scalar theory where two additional fields with 
masses $M_1=m+\tilde\Lambda$ and $M_2=m+2\tilde\Lambda$ and coefficients
$C_1=-2$ and $C_2=1$ are sufficient to make the theory finite.
In the case of a non linearly realized symmetry (as in this case for the 
pseudoscalar sector) quartic divergences due to the non invariance
of the measure have to be treated before.

\vfill\eject

\end{document}